\documentclass[orivec]{llncs}
\usepackage{latexsym}
\usepackage{amsmath}
\usepackage{graphicx}
\begin{document}

\newcommand{\vect}[1]{\overrightarrow{#1}}

\title{Statistical Automatic Summarization\\ in Organic Chemistry}

\titlerunning{A Chemical automatic Summarizer}

\author{Florian Boudin\inst{2}, Patricia Vel\'azquez-Morales and Juan-Manuel Torres-Moreno\inst{1}}

\authorrunning{Florian Boudin et al.} 

\institute{Laboratoire Informatique d'Avignon \\
339 chemin des Meinajari\`es, BP1228 \\
84911 Avignon Cedex 9, France. \\
\email{juan-manuel.torres@univ-avignon.fr }  \vspace*{0.5cm}
\and Universit\'e de Montr\'eal - RALI\\
C.P. 6128, succursale Centre-ville \\
CP 6079 Succ. Centre Ville H3C 3J7  \\
Montr\'eal (Qu\'ebec), Canada. \\
\email{boudinfl@iro.umontreal.ca  }  
}

\date{}

\maketitle

\begin{abstract} 
We present an oriented numerical summarizer algorithm, applied to producing automatic summaries of scientific documents in Organic Chemistry.
We present its implementation named \textsc{Yachs} (Yet Another Chemistry Summarizer) that combines a specific document pre-processing with a sentence scoring method relying on the statistical properties of documents.
We show that \textsc{Yachs} achieves the best results among several other summarizers on a corpus of Organic Chemistry articles. 
\end{abstract} 

\section{Introduction}
Over 1.7 million new Chemistry articles were published in 2007\footnote{See Chemical Abstracts Publication Record, \textit{http://www.cas.org/}}, thereby most of scientists today are on \textit{information overload}.
Information extraction technology arose in response to the need for efficient processing of documents in specialized domains.
Scientists, especially chemists, want to be able to promptly access information concealed in a document in addition to the author's abstract that is often too concise or not satisfying.
Automatically producing summaries from Organic Chemistry documents is a challenging but critical task for chemical information retrieval.
\textit{Text Summarization} is the process of distilling the most important information from a source (or sources) to produce an abridged version for a particular user and task \cite{mani1999}.
There are many uses of text summarization in everyday activities, we are familiar with summaries such as headlines, reviews or digests.
Introduced by Luhn \cite{luhn1958} in the late 1950's, text summarization was characterized by the use of a surface level approach (i.e. exploiting term frequencies).
The first entity-level approaches based on syntactic analysis appeared in the early 1960's \cite{climenson1961} while the use of location features and cue phrases was not developed until later \cite{edmundson1969}.
The investigations reported by \cite{pollock1975} at the Chemical Abstracts Service (CAS) provide further insight into the effectiveness of automatic summarization in particular domain areas.
Corpus-based approaches were introduced by \cite{kupiec1995} with a trainable summarization system using a collection of text/summaries pairs as train-set.
A Bayes classifier algorithm takes each sentence and, based on features such as cue phrases, sentence length or location, computes a probability that it should be included in the summary.
Thereafter, \cite{mani1998} have extended this model using decision tree rules instead of bayesian classifiers.
Rhetorical status was proposed by \cite{teufel2002} to summarize scientific articles (Computational Linguistic conference articles) that can highlight the new contribution of the source article.
The limitations of this approach is that it depends on manual resources (metadiscourse features are manually annotated).
\cite{reeve2007} proposes to combine semantic-based and frequency-distribution approaches for extractive text summarization in biomedical documents.
However, this approach requires a difficult concept identification process.
Benefits of automatic abstracting are now clearly identified: it is inexpensive compared to human effort and, unlike humans, it is consistent and avoid subjectivity and variability observed in human abstractors.
Typically, summarization systems are two-phased, consisting of a content selection step followed by a generation step.
Firstly, text fragments (most often sentences) are assigned a score that reflects how important they are.
The highest-ranking material can then be arranged and displayed as ``extracts".
This paper presents \textsc{yachs} (Yet Another Chemistry Summarizer), a summarization system that generates extracts from scientific articles in specialized domain, Organic Chemistry.
The motivation behind this work is to allow non-experts users to access information contained in high-end scientific documents by dynamically generating extracts.
Specifically, through statistical entity level approaches, we seek to produce highly informative extracts that can stand in place of the original author's abstracts as surrogates.

\section{Method}
\label{sec:sys}

\subsection{Pre-processing}
\label{sec:pre}
The first question we are concerned with is whether classical Natural Language Processing (NLP) tools are consistent within the organic chemistry domain. The answer is clearly no.
Tools such as parsers, taggers or chunkers achieve very poor on these documents without requiring a strenuous, costly and often manual adaptation phase. 
Issues encountered by classical tools are due to domain specificity: very wide vocabulary, long sentences containing \textit{noise} (i.e. citations, chemical formulas, tables, pictures references, etc.), high quantity of \textit{hapax legomena}\footnote{Terms which only appears once in a document.}, etc.

The basic idea is to represent the document within the vector space model introduced by \cite{salton1975} and apply specific numeric treatments to select the most salient sentences.
An $n$-dimensional term-space $\varGamma$, where $n$ is the number of different terms found in the document, is constructed.
One convenient way to represent the document in $\varGamma$ is a matrix $M = [a_{x,y}]_{\ x=1 \dots m;\ y=1 \dots n}$ where $m$ is the number of sentences and $n$ the number of different terms.
In this interpretation, every row of $M$ is a vector $\vec{s}_x$ representing the sentence $x$ in which each component is the term frequency within the sentence.

In order to reduce the size of the matrix $M$ and accordingly reduce the computational complexity, some reductions and filtering are applied to sentences (see table \ref{exemple_processing}).
In written language, some words carry more \textit{meaning} than others.
Thereby, a stop-words elimination phase is performed (1) to delete non representative words (i.e. words such as `\textit{the}', `\textit{of}', `\textit{in}'... are removed).
One standard pre-processing would normalize character case, remove punctuation and special characters (2).

However, important information about chemical compounds may be lost during the filtering process (e.g. `1,2-$dienes$' is transformed into `$dienes$').
Besides if word normalization (i.e. stemming \footnote{The porter stemmer algorithm \cite{porter1980} is used to normalize words by removing commoner morphological and inflexional endings from words.}) is applied afterwards (3), erroneous information is brought in the sentence (e.g. `1,2-$dienes$' is transformed into `$dien$').
We propose to perform a chemical compounds detection to protect these terms during the normalization process (2$^\prime$).
Finally stemming is performed only on non-chemical terms (3$^\prime$).
%
%
\begin{table}[ht!]
\begin{center}
\begin{tabular}{l p{0.88\textwidth}}
\hline
\textbf{Original} & Cycloalkynes are known to isomerize to the 1,2-dienes under basic conditions.\\ \hline
\textbf{(1)} & Cycloalkynes known isomerize 1,2-dienes under basic conditions.\\
\textbf{(2)} & cycloalkynes known isomerize dienes under basic conditions\\
\vspace{0.1cm}
\textbf{(3)} &  cycloalkyn know isomer dien under basic condit\\
\textbf{(2$^\prime$)} &  \textbf{cycloalkynes} know isomerize \textbf{1,2-dienes} under basic conditions\\
\textbf{(3$^\prime$)} &  \textbf{cycloalkynes} \textit{know} \textit{isomer} \textbf{1,2-dienes} \textit{under} \textit{basic} \textit{condit}\\
\hline
\end{tabular}
\end{center}
\caption{Example of sentence pre-processing.}
\label{exemple_processing}
\end{table}

Chemical compounds are detected within sentences using a combination of two classifiers.
The first one is a Bayes classifier trained on 3-grams of letters whereas the second one uses pattern matching with a small number of manually written rules (7 rules).
Each sentence is tokenized in words and each word is classified by the two classifiers, precision is prioritized by using the \textsc{and} combination (i.e. a word has to be classified as chemical compound by the two classifiers).
This hybrid approach (statistical and symbolic) for chemical term recognition achieves very good results on a test corpus composed by Organic Chemistry articles \cite{boudin2008}.

\subsection{Sentence Ranking}
Once sentences are pre-processed, a combination of features (also called metrics) is used to assign a score to each sentence.
That score reflects how important the sentences are in relation to the whole document.
The main advantage of this approach is that \textit{zero knowledge} is required and that makes the system fully adjustable to any language and/or domain.
This section formally describe the metrics calculated by \textsc{yachs}.

Authors normally conceive titles as circumscribing the topic of the document.
Sentences sharing words, containing words related to or similar with the title are likely to be relevant.
Following this assumption, two metrics computing similarity measures between a sentence and the title have been implemented.
The first measure is the well known cosine angle \cite{salton1975} between a sentence and the title vectorial representations in $\varGamma$.
The main weakness of $cosine$ and more generally of all similarity measures using words for tokens is that they are relying too much on term normalization.
Their performance dramatically decrease with wrongly or non normalized words.
We propose a second similarity measure based on the Jaro-Winkler distance \cite{winkler1999} that can bridge morphologically similar words in order to smooth normalization and misspelling errors.
The original Jaro-Winkler measure, denoted \textsc{Jw}, uses the number of matching characters and transpositions to compute a similarity score between two terms, giving more favourable ratings to terms that match from the beginning (see examples in table \ref{table:exjwmots}).
We have extended this measure to calculate the similarity between a sentence $s_m$ and the title $t$ (see table \ref{exemple_similarity}):

\begin{equation}
\textsc{Jw}_e(s_x,t) = \frac{1}{|t|} \cdot \sum_{w_t \in t} \max_{w_x \in S^\prime}  \textsc{Jw}(w_t,w_x)
\label{eq:jaro}
\end{equation}

\noindent where $S^\prime$ is the term set of $s_x$ in which the words $w_x$ that already have maximized $\textsc{Jw}(w_t,w_x)$ are removed.

\begin{table}[ht!]
\begin{center}
\begin{tabular}{llr}
\hline
\textbf{Word 1} & \textbf{Word 2} & \textsc{Jw} \\
\hline
nucleophile & nucleophilic & 0.94515 \\
nucleophile & electrophile & 0.47643 \\
diphenyl & 1,1-Diphenylmethanone\ \ & 0.35516 \\
1,1-Diphenylmethanone\ \ & nucleophile & 0.11038 \\
\hline
\end{tabular}
\end{center}
\caption{Examples of Jaro-Winkler distance (\textsc{Jw}) between words.}
\label{table:exjwmots}
\end{table}%
%
\begin{table}[ht!]
\begin{center}
\begin{tabular}{l p{0.87\textwidth}}
\hline
\textbf{Title} & {\scriptsize Generation of Cycloalkynes by Hydro-Iodonio-Elimination of Vinyl Iodonium Salts}\\
\textbf{Sentence} & {\scriptsize Cycloalkylidenecarbene can provide a ring-expanded cycloalkyne via 1,2-rearrangement.}\\
\hline
\textbf{T}$_{preproc.}$ & \textit{generat} \textbf{cycloalkynes} \textit{hydro-lodonio-elimination} \textbf{vinyl iodonium salt}\\
\textbf{S}$_{preproc.}$ & \textbf{cycloalkylidenecarbene} \textit{provid ring expand} \textbf{cycloalkyne} \textit{via rearrang}\\
$cosine$ &  0 (no co-occurrencies) \\
$\textsc{Jw}_e$ &  0.43348 \\
\hline
\end{tabular}
\end{center}
\caption{Example of similarity measures between the title and a sentence (\textbf{T}$_{preproc.}$ and \textbf{S}$_{preproc.}$ are the pre-processed title and the pre-processed sentence). }
\label{exemple_similarity}
\end{table}

Experiments have shown that sentence position within the document is a very important feature \cite{mani1999}.
Indeed, the information is not homogeneously spread across the document but scattered tidily by the author respecting universally accepted writing rules.
Document beginnings and endings usually contain sentences that are highly relevant because their original goals are to present and sum up the topic.
Sentence position is therefore used as metric, denoted $P$ (equation \ref{P}), by computing a smoothing parabola depending on the number of sentence $m$ in the document.

\begin{equation}
  P_{x} =
     \begin{cases}
        \frac{(x-1) - (\frac{m}{2} - 1)}{\frac{m}{2}} & \text{if $m$ is even} \\
        \frac{(x-1) - \frac{m}{2}}{\frac{m}{2}} & \text{Otherwise}
     \end{cases}
     \label{P}
\end{equation}

We have implemented four other metrics relying on numerical treatments, they are computed on the matrix $M$ (previously introduced in section \ref{sec:pre}).
The first one is the sum of word frequencies, denoted $F$ (equation \ref{F}), that uses the frequencies of words in sentences.
Sentences that are containing \textit{important} words are considered as relevant.
\begin{equation}
F_{x} = \sum_{y = 1}^{n} a_{x,y} \label{F} 
\end{equation}

The second metric, denoted $C$ (equation \ref{C}), relies on the number of chemical compounds detected in the sentence giving a penalty to sentences that do not contain any chemical compounds.
\begin{equation}
C_{x} = \begin{cases}
        1 & \text{if $x$ contains at least one chemical compound} \\
        0 & \text{Otherwise}
     \end{cases}
 \label{C} 
\end{equation}

The third metric, denoted $I$ (equation \ref{I}), represents the interaction relationship between sentences.
The underlying idea is that sentences containing words that are used in other sentences are statistically more representative for the document \cite{torres2002}.
\begin{equation}
I_{x} = \sum_{{\scriptstyle y=1} \atop \scriptstyle a_{x,y} \neq 0}^{n} \sum_{{\scriptstyle z = 1} \atop \scriptstyle z \neq x}^{m} a_{z,y}\label{I}
\end{equation}

The last metric, denoted $H$ (equation \ref{H}), is the sum of the Hamming distances computed on the sentence pair words \cite{torres2002}.
The idea is to give more weight to pairs of words that appears independently in sentences.
Synonyms and topic-related words generally are, according to the Hamming distance, high weighted.
In order to compute this metric, a second matrix denoted $M_h$ is constructed from $M$.
$M_h$ is a $n \times n$ triangular matrix constructed from word co-occurrences between sentence pairs:
\begin{align}
	M_h & = [h_{i,j}]_{\ i=1 \dots n;\ j=1 \dots n} \nonumber \\ 
	h_{i,j} & = \sum_{x = 0}^{m}
		\begin{cases}
			1 &  \text{\ \ \ if $a_{x,i} \neq a_{x,j}$} \\
			 0 & \text{\ \ \ Otherwise}
		\end{cases}
		\nonumber \\
	H_{x} & = \sum_{i = 1}^{n-1} \sum_{j = i+1}^{n}
			\begin{cases}
			h_{i,j} &  \text{\ \ \ if $a_{x,i} \neq  0$ and $a_{x,j}  \neq 0$} \\
			0 & \text{\ \ \ Otherwise}
		\end{cases}
	\label{H}
\end{align}

Sentences are scored by using a equiprobable linear combination\footnote{Other combinations might be considered, but a large training corpus is required to tuned the parameters.} of the normalized metrics (i.e. ranged in $[0,1]$) described above.
A ranked sentences list is produced by the system allowing to construct the extract by arranging the high scored sentences until the desired size is reached.

\section{Experimental Settings}
\label{sec:experimental}
Considerable interest has been expressed and effort expended in attempting to evaluate automatically the quality of the summaries.
There exists two different types of evaluation: extrinsic and intrinsic \cite{jones1996}.
Extrinsic evaluations measure the quality of a summary based on how it affects certain tasks.
In intrinsic evaluations, summary's quality is evaluated by an analysis of its content.
Most existing automated evaluation methods work by comparing the produced summaries to one or more reference summaries (ideally, produced by humans).
In order to evaluate our system, we have collected a testing set from \textit{http://pubs.acs.org}.
The testing set is composed by 100 pairs of articles/abstracts coming from different journals (Organic Letters, Accounts of Chemical Research and Journal of Organic Chemistry) of different years (respectively 2000-2002, 2005-2007 and 2007-2008), different authors and topics.
Each document has been cleaned up manually from the PDF (or HTML) version (figures, bibliographic references, special characters, etc. have been removed).
By ways of comparison the corpus used in the Document Understanding Conference (DUC)\footnote{Document Understanding Conferences are competitions on text summarization conducted since 2000 by the National Institute of Standards and Technology (NIST), \textit{http://www-nlpir.nist.gov}} 2005 competition was also composed of 100 sets. 
Table \ref{table:composition} shows some statistics about the testing set.

\begin{table}[htdp]

\begin{center}
\begin{tabular}{lcccc}
\hline
\textbf{Journal} & \textbf{Year} & \textbf{Number} & \textbf{Sentences} & \textbf{Words}\\ \hline
Organic Letters & 2000-2008 & 63 &  5.313 & 104.588\\ 
Accounts of Chemical Research  & 2005-2006 & 10 & 979 & 18.337 \\ 
The Journal of Organic Chemistry\ \ & 2007-2008 & 27 & 2.631 & 66.242\\ \hline
\textbf{Total} & - & 100 & 8.923 & 189.167 \\ \hline
\end{tabular}
\end{center}
\caption{Testing corpus description.}
\label{table:composition}
\end{table}

\subsection{Performance Measures}
To evaluate the quality of our generated summaries, we choose to use the \textsc{Rouge}\footnote{\textsc{Rouge} is available at \textit{http://haydn.isi.edu/ROUGE/}.} \cite{lin2004} evaluation toolkit, that has been found to be highly correlated with human judgments \cite{dang2005}.
\textsc{Rouge-n} is a $n$-gram recall measure calculated between a candidate summary and one or more reference summaries.
In our experiments \textsc{Rouge-1}, \textsc{Rouge-2} and \textsc{Rouge-su4} will be computed.
Each generated extract will be evaluated by comparison with the author's abstract.
The size of the produced extracts is set at 5\% of the original document (in sentence number) with a minimum of three sentences.

\section{Results}
\begin{figure}[!ht]
	\centering
	\includegraphics[width=12.2cm]{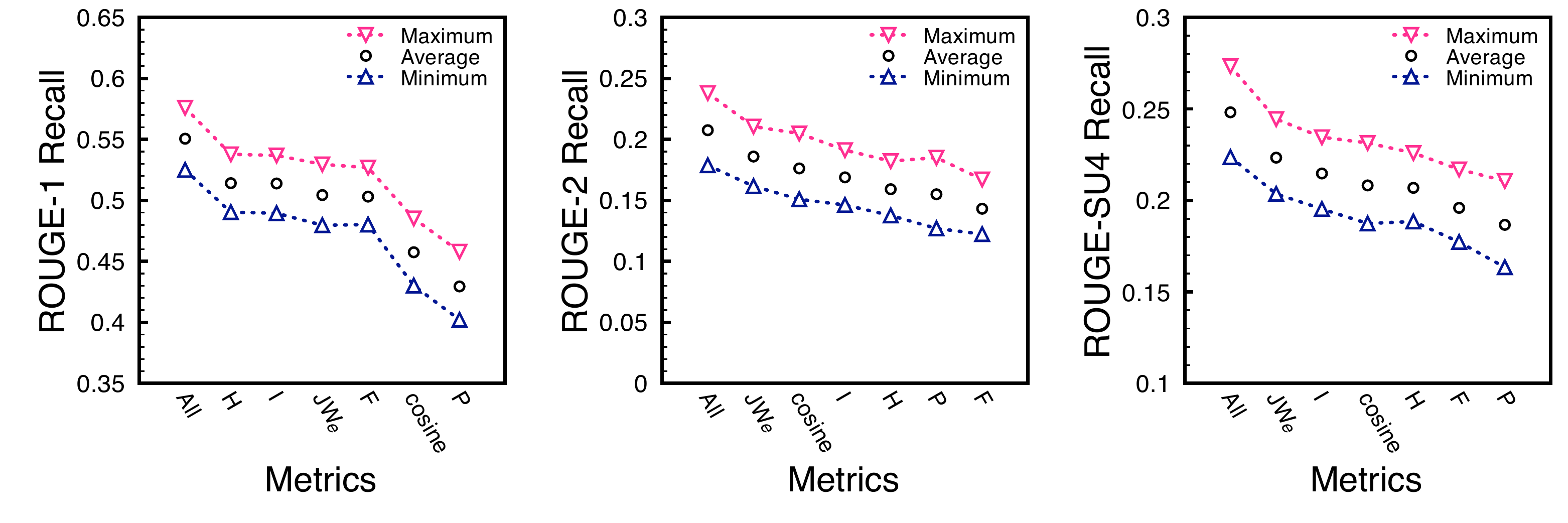}
	\caption{\textsc{Rouge-1}, \textsc{Rouge-2} and \textsc{Rouge-su4} recall scores for each metric independently and for their combination (denoted  {\scriptsize All}).}
	\label{fig:metriques}
\end{figure}

The first experiment is focused on the study of metrics. Figure \ref{fig:metriques} shows the \textsc{Rouge} results of each metric alone and their combination.
As we can see from these results, the combination, denoted by {\scriptsize All}, always outperforms the best metric alone.
The most discriminant metrics are the similarity measures with the title ($\textsc{Jw}_e$ and $cosine$) and the interaction relationship between sentences ($I$).
The title similarity measures allow to focus the summary on the document main topic, delineated by the author.
The similarity measure $\textsc{Jw}_e$ that we propose is globally the most discriminant metric, its ability to bridge morphologically similar words is well adapted for Organic Chemistry documents.
The interaction metric uses the networks built by words within the document to compute a relevance score, sentences that are constructed with terms appearing in many other sentences are selected.
These sentences are judged as being the most representative to the document because they are containing most of the information.

\begin{figure}[!ht]
	\centering
	\includegraphics[width=12.2cm]{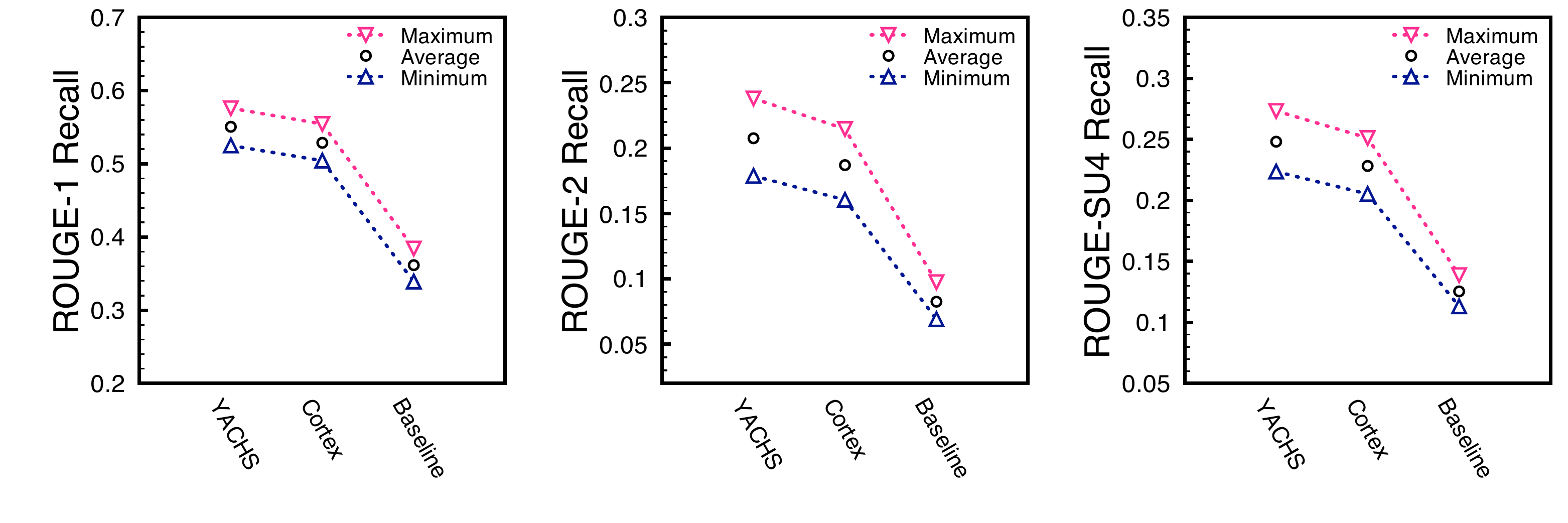}
	\caption{\textsc{Rouge-1}, \textsc{Rouge-2} and \textsc{Rouge-su4} recall scores of  \textsc{yachs}, Cortex and the random baseline.}
	\label{fig:system_human}
\end{figure}

A second evaluation compares \textsc{yachs} to a generic statistical summarizer and a baseline on the corpus of manually segmented documents (see Figure \ref{fig:system_human}).
We use the Cortex summarizer \cite{torres2002} which is based on the same approach that \textsc{yachs}, namely a combination of relevance metrics, but without the chemical compounds detection process and the powerful $\textsc{Jw}_e$ metric.
The baseline is generated by arranging $n$ sentences selected randomly from the document, $n$ being 5\% of the document sentence number with a minimum of three sentences.
In order to smooth the baseline results, the average of 100 baseline evaluations is used in our experiments.
\textsc{yachs} achieves the best results among the \textsc{Rouge} evaluations.
It confirms that the specialized pre-processing and sentence scoring are well adapted to process domain specialized (Organic Chemistry) documents.

\begin{figure}[!ht]
	\centering
	\includegraphics[width=12.2cm]{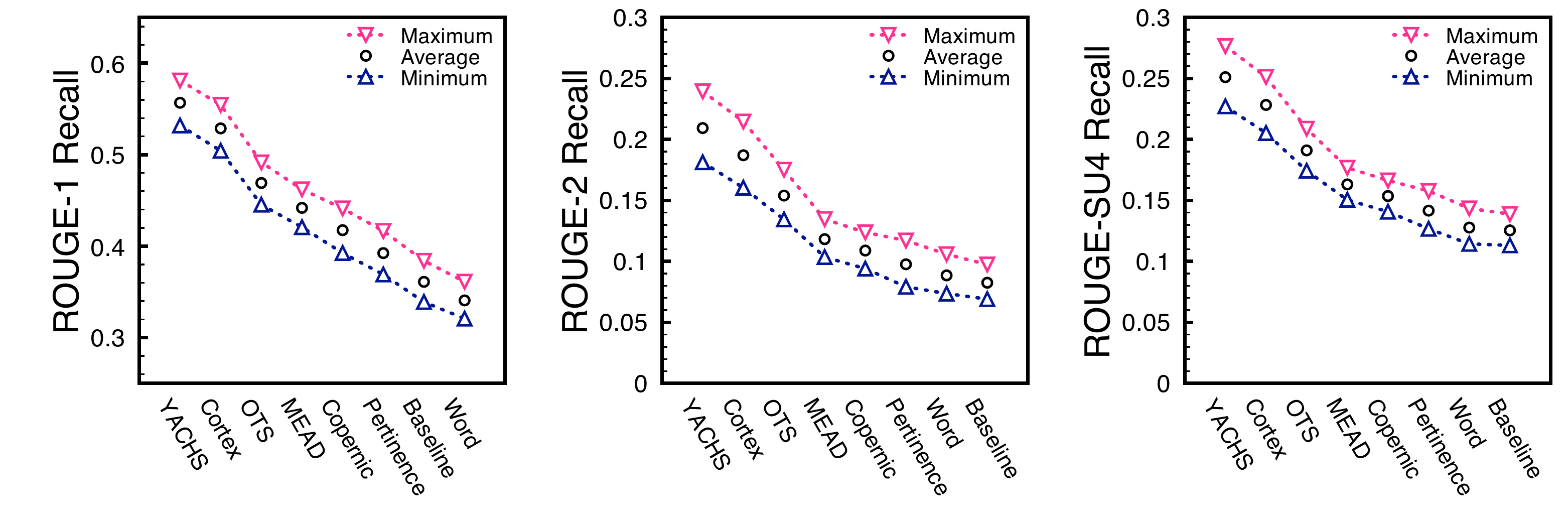}
	\caption{Comparison of the \textsc{Rouge-1}, \textsc{Rouge-2} and \textsc{Rouge-su4} recall scores for the seven summarizers and the random baseline.}
	\label{fig:systems}
\end{figure}

The last evaluation models a real world summarization task: a plain text is given as input (without manual sentence segmentation), each summarizer has to produce an extract of size equals to 5\% of the original document (in sentence number).
We compare \textsc{yachs} to six extractive summarizers and one baseline, results are shown in figure \ref{fig:systems}.
\textsc{yachs}, Cortex and the baseline use the same automatic sentence segmentation process which consists in a standard sentence boundaries detection system enriched with lists of abbreviations.
The other systems using their own sentence splitters.
The baseline is generated by arranging $n$ sentences selected randomly from the document, $n$ being 5\% of the document sentence number with a minimum of three sentences.
Again, the average of 100 baseline evaluations is used in our experiments.
MEAD\footnote{Available at \textit{http://www.summarization.com/mead/}} is a centroid based summarizer \cite{radev2001} that extract sentences according to three features: sentence centrality within the cluster, sentence position within the document and weighted similarity with the title.
Open Text Summarizer (OTS) \cite{yatsko2007} is an Open Source project that, similarly to MEAD, use statistical word-frequency methods to score sentences that are beforehand parsed.
It also incorporates an English language lexicon with synonyms and cue terms.
Pertinence Summarizer\footnote{Available at \textit{http://www.pertinence.net/ps/}} performs linguistic processing of a document to generates an extract, the sentence scoring method considering general and specialized (Chemistry) linguistic markers.
Besides, two frequency-based summarizers are evaluated: Copernic\footnote{Available at \textit{http://www.copernic.com/en/products/summarizer/index.html}} summarizer and the AutoSummarize feature of Microsoft Word.
Exact details of their algorithms are alas not documented.

\textsc{yachs} and Cortex clearly stand out from the crowd.
A significant score margin separates these two systems with the others confirming that these statistical techniques work well for Organic Chemistry documents.
\textsc{yachs} achieves the best results among all summarizers proving that specialized pre-processing and adapted sentence scoring are features allowing to generate better specialized extracts.

\section{Conclusion}
In this paper we have described an efficient approach for automatically generating extracts from documents in Organic Chemistry.
Through experiments performed on a corpus composed of scientific articles, we have showed that our approach (implemented in the \textsc{yachs}\footnote{An demonstration version of \textsc{yachs} is available at \textit{http://daniel.iut.univ-metz.fr/yachs}} system) achieves promising results.
This work represent a good starting point but do show a critical point: a lot of information is lost during document pre-processing.
Indeed, pictures, tables or captions, that are removed during PDF (or HTML) to text conversion, are containing salient information that can be used to enhance extracts.
Among the others, there are several points that would be worthy of further investigation:
\begin{itemize}
\item Use multi-media information (i.e. pictures, texts, tables, etc.) to generate extracts.
\item Fuse text summarization and Question Answering (QA) to model real-world complex QA, in which a question cannot be answered by simply stating a name, date, quantity, etc.
\end{itemize}


\bibliographystyle{splncs}
\bibliography{sum_chem}

\begin{thebibliography}{10}

\bibitem{mani1999}
Mani, I., Maybury, M.T.:
\newblock {Advances in Automatic Text Summarization}.
\newblock The MIT Press (1999)

\bibitem{luhn1958}
Luhn, H.P.:
\newblock {The Automatic Creation of Literature Abstracts}.
\newblock IBM Journal of Research and Development \textbf{2}(2) (1958)  159

\bibitem{climenson1961}
Climenson, W.D., Hardwick, N.H., Jacobson, S.N.:
\newblock {Automatic syntax analysis in machine indexing and abstracting}.
\newblock American Documentation \textbf{12}(3) (1961)  178--183

\bibitem{edmundson1969}
Edmundson, H.P.:
\newblock {New Methods in Automatic Extracting}.
\newblock Journal of the ACM (JACM) \textbf{16}(2) (1969)  264--285

\bibitem{pollock1975}
Pollock, J.J., Zamora, A.:
\newblock {Automatic Abstracting Research at Chemical Abstracts Service}.
\newblock Journal of Chemical Information and Computer Sciences \textbf{15}(4)
  (1975)  226--232

\bibitem{kupiec1995}
Kupiec, J., Pedersen, J., Chen, F.:
\newblock {A trainable document summarizer}.
\newblock In: 18th annual international ACM SIGIR conference on Research and
  development in information retrieval, ACM Press New York, NY, USA (1995)
  68--73

\bibitem{mani1998}
Mani, I., Bloedorn, E.:
\newblock {Machine learning of generic and user-focused summarization}.
\newblock In: 15th National Conference on Artificial intelligence (AAAI), AAAI
  press, Menlo Park, CA, USA (1998)  820--826

\bibitem{teufel2002}
Teufel, S., Moens, M.:
\newblock {Summarizing scientific articles: experiments with relevance and
  rhetorical status}.
\newblock Computational Linguistics \textbf{28}(4) (2002)  409--445

\bibitem{reeve2007}
Reeve, L.H., Han, H., Brooks, A.D.:
\newblock {The use of domain-specific concepts in biomedical text
  summarization}.
\newblock Information Processing and Management \textbf{43}(6) (2007)
  1765--1776

\bibitem{salton1975}
Salton, G., Wong, A., Yang, C.S.:
\newblock {A Vector Space Model for Automatic Indexing}.
\newblock Communications of the ACM \textbf{18}(11) (1975)  613--620

\bibitem{porter1980}
Porter, M.F.:
\newblock {An Algorithm for Suffix Stripping}.
\newblock Program \textbf{14} (1980)  130--137

\bibitem{boudin2008}
Boudin, F., Torres-Moreno, J.M.:
\newblock {Mixing Statistical and Symbolic Approaches for Chemical Names
  Recognition}.
\newblock In: Conference on Intelligent Text Processing and Computational
  Linguistics (CICLing). Volume 4919., Springer (2008)  334--349

\bibitem{winkler1999}
Winkler, W.E.:
\newblock {The state of record linkage and current research problems}.
\newblock Statistics of Income Division \textbf{4} (1999)  73--79

\bibitem{torres2002}
Torres-Moreno, J.M., Velazquez-Morales, P., Meunier, J.G.:
\newblock {Condens{\'e}s de textes par des m{\'e}thodes num{\'e}riques}.
\newblock In: JADT. Volume~2. (2002)  723--734

\bibitem{jones1996}
Sp\"{a}rck~Jones, K., Galliers, J.R.:
\newblock {Evaluating Natural Language Processing Systems: An Analysis and
  Review}.
\newblock Springer (1996)

\bibitem{lin2004}
Lin, C.Y.:
\newblock {Rouge: A Package for Automatic Evaluation of Summaries}.
\newblock In: Workshop on Text Summarization Branches Out. (2004)  25--26

\bibitem{dang2005}
Dang, H.T.:
\newblock {Overview of DUC 2005}.
\newblock In: DUC. (2005)

\bibitem{radev2001}
Radev, D.R., Blair-Goldensohn, S., Zhang, Z.:
\newblock {Experiments in Single and Multi-Document Summarization Using MEAD}.
\newblock In: DUC. (2001)

\bibitem{yatsko2007}
Yatsko, V.A., Vishnyakov, T.N.:
\newblock {A method for evaluating modern systems of automatic text
  summarization}.
\newblock Automatic Documentation and Mathematical Linguistics \textbf{41}(3)
  (2007)  93--103

\end{thebibliography}

\end{document}